# A Comparison of SC-FDE and UW DFT-s-OFDM for Millimeter Wave Communications

Alphan Şahin, Rui Yang, Frank La Sita, Robert L. Olesen
{Alphan.Sahin, Rui.Yang, Frank.LaSita, Robert.Olesen}@InterDigital.com

*Abstract*—In this study, we compare the single-carrier (SC) waveform adopted in IEEE 802.11ad and unique word discrete Fourier transform spread orthogonal frequency division multiplexing (UW DFT-s-OFDM) waveform. We provide equivalent representations of up-sampling and down-sampling operations of the SC waveform by using discrete Fourier transform (DFT) and inverse DFT to enable explicit comparison of these two similar waveforms. By using this representation, we discuss why the IEEE 802.11ad SC waveform can cause suboptimal performance in multipath channel and discuss how to improve it with UW DFT-s-OFDM. With comprehensive link-level simulations, we show that replacing the 802.11ad SC waveform with UW DFT-spread OFDM can result in 1 dB gain in peak throughput without affecting the IEEE 802.11ad packet structure. We also evaluate the cross links where the transmitter is UW-DFT-s-OFDM and the receiver is traditional SC-FDE or vice versa. We demonstrate that UW DFT-s-OFDM receiver can decode an IEEE 802.11ad SC waveform with a slight SNR loss while IEEE 802.11ad SC receiver can decode a UW DFT-spread OFDM waveform with an interference floor.

## I. INTRODUCTION

Millimeter wave (mmWave) communication is considered one of the key components of the Fifth Generation (5G) cellular and Wi-Fi networks. With the availability of wide bandwidth at higher gigahertz frequencies, mmWave communications can achieve tens of gigabits per second data throughputs [1]. However, radio communications in high frequency bands may suffer from several physical limitations such as severe path loss and penetration loss. With narrow analog beams using advanced beamforming technologies, the coverage range may be extended depending on the maximum transmit power level. Hence, the energy efficiency has become a key design criterion for the mmWave communication networks.

At the physical layer, a waveform not only needs to enable data to be transmitted with high spectral efficiency, but also needs to address hardware challenges such as non-linear distortion from the power amplifier (PA) and energy efficiency. A waveform that has high peak-to-average power ratio (PAPR) requires a large amount of power back-off at the input to the PA to avoid non-linear signal distortion, which in turn reduces PA efficiency and coverage area. Therefore, single carrier (SC) schemes, which have a PAPR advantage over multicarrier schemes, are some of the most appealing waveforms for mmWave communications.

In the literature, there have been numerous studies that discuss SC waveforms, particularly from the perspective of frequency domain (FD) processing techniques at receiver (e.g., [2], [3] and the references therein). For SC waveforms, low-complexity FD processing techniques are typically enabled by the methods that convert the linear convolution of the channel to a circular convolution. For example, extending the data symbol block with a sequence (i.e., known as unique word (UW) [2], pseudo noise extension [3], or training sequence [4]), or cyclically extending the data blocks before pulse shaping are the traditional approaches that convert the impact of channel on data blocks to a circular convolution at the receiver for SC systems. Among these methods, it has been shown that extending the data block with a sequence can be more beneficial than the cyclic extension as it enables various non-linear FD equalization techniques [3], time-frequency synchronization [4], channel estimation, phase tracking, and noise estimation [5] for SC waveforms. Due to these advantages, SC with UW, adopted in IEEE 802.11ad [6], is a promising waveform for mmWave communications. It is worth noting that the concept of UW is also applied to OFDM at the expense of higher transmitter and receiver complexity as compared to cyclic prefix (CP) OFDM [7].

Historically, a major innovation for SC systems occurred through the introduction of a DFT precoder in FD to an OFDM waveform [8] [9], known as cyclic prefix discrete Fourier transform spread orthogonal frequency division multiplexing (CP DFT-s-OFDM). One can show that the data symbols are convolved with a Dirichlet sinc function circularly in time with CP DFT-s-OFDM [10], which is similar to SC waveforms. However, since CP DFT-s-OFDM is also a precoded OFDM symbol, it enables efficient frequency domain multiple access (FDMA) by eliminating the guard bands required for the one with traditional SC waveforms. Due to its PAPR advantage compared to OFDM and its excellent efficieny in FDMA, it was adopted for the use in the LTE uplink [11] (The reader is also refered to the discussions on CP DFT-s-OFDM and OFDM in [12]). Since its adoption, there have been various notable studies on DFT-s-OFDM and its variants. For example, the space-frequency block code (SFBC) scheme which preserves the PAPR advantage of DFT-s-OFDM is proposed in [13]. In [14] [15] [16], the CP extension of CP DFT-s-OFDM is eliminated by padding zeros to the beginning and the last part of the data block. In [17] and [18], a scheme which is called UW DFT-s-OFDM is proposed by showing that the zeros can be replaced by a set of fixed symbols, i.e., UW sequence, as done in the traditional SC waveform. In this scheme, the last parts of the UW DFT-s-OFDM symbols, i.e., tails, are *approximately* the same. After the UW DFT-s-OFDM symbols are concatenated in time, the tail part of the $i$th symbol serves as the CP for the $(i + 1)$th UW

DFT-s-OFDM symbol. Hence, no additional CP is required for the $(i + 1)$th UW DFT-s-OFDM symbol. In addition, various cancellation signals are proposed to improve the tail characteristics of UW DFT-s-OFDM by mitigating the leakage from the data symbols at the tail in [17] and [18]. In [10], a low-complexity suppression is proposed by applying a windowing to the output of DFT block in FD.

In this paper, in contrast to previous studies on SC and UW DFT-s-OFDM, we compare the SC waveform adopted in IEEE 802.11ad [6] and UW DFT-s-OFDM by expressing equivalent operations for the up-sampling and down-sampling blocks of the SC transmitter and receiver by using DFT and inverse DFT. By using this representation, we show the followings:
1) The synthesis of the SC waveform is shown using block-based operations, indicative of the compatibility of the numerology for UW DFT-s-OFDM to the numerology of the SC adopted in IEEE 802.11ad. Hence, UW DFT-s-OFDM does not change the packet structure which in turn leads to a framework without less standardization effort.
2) The down-sampling at the SC receiver causes *incoherent* combinations in the FD under a fading channel, which causes degradation at the receiver. By providing peak throughput, we show that UW DFT-s-OFDM provides 1 dB gain as compared to the SC waveform.
3) We discuss FD windowing in [10] for UW DFT-s-OFDM and develop a two-stage equalization to address the non-coherent additions to improve error rate performance.
4) We demonstrate that the UW DFT-s-OFDM receiver can decode the IEEE 802.11ad SC waveform with a slight SNR loss while the IEEE 802.11ad SC receiver can decode UW DFT-s-OFDM with an interference floor.

The rest of the paper is organized as follows. In Section II, we provide system models for SC and UW DFT-s-OFDM. In Section III, we discuss the representation of the SC transmitter and receiver while comparing them to that for UW DFT-s-OFDM. In Section IV, we present numerical results performed under IEEE 802.11ad framework. We conclude the paper in Section V with final remarks.

Notations: Matrices [columns vectors] are denoted with upper [lower] case boldface letters (e.g., $\mathbf{A}$ and $[\mathbf{a}]$). The Hermitian and the transpose operations are denoted by $(\cdot)^H$ and $(\cdot)^T$, respectively. The conjugate of $\mathbf{a}$ is denoted by $\bar{\mathbf{a}}$. The symbols $*$ and $\otimes$ denote linear convolution and Kronecker product, respectively. The $\ell_2$-norm of $\mathbf{a}$ is denoted by $\|\mathbf{a}\|_2$. The operation flip$\{\mathbf{a}\}$, the operation circ$\{\mathbf{a}, \tau\}$, and the operation diag$\{\mathbf{a}\}$ reverse the order of the elements of $\mathbf{a}$, circularly shift $\mathbf{a}$ by $\tau$, and create a diagonal matrix where its diagonal is $\mathbf{a}$, respectively. The set of complex, real, and integer numbers are shown as $\mathbb{C}$, $\mathbb{R}$, and $\mathbb{Z}$, respectively. $\mathbb{Z}^+$ symbolizes the set of positive integers. The multivariate complex Gaussian distribution is denoted by $\mathcal{CN}(\boldsymbol{\mu}, \mathbf{C})$, where $\boldsymbol{\mu}$ is the mean vector and $\mathbf{C}$ is the covariance matrix. $\mathbf{I}_N$, $\mathbf{0}_{N\times M}$, $\mathbf{1}_{N\times M}$ are the $N\times N$ identity, $N\times M$ zero, and $N\times M$ one matrices, respectively.

## II. SYSTEM MODEL

### A. Single Carrier Waveform

We consider an SC packet based on IEEE 802.11ad [6] which consists of multiple SC blocks. Each SC block is generated from a group of data and fixed symbols via up-sampling, filtering, and down-sampling operations. Let data and fixed symbols transmitted within the $i$th SC block be the elements of vector $\mathbf{d}_i \in \mathbb{C}^{M_d \times 1}$ and $\mathbf{s} \in \mathbb{C}^{M_s \times 1}$, respectively, where $M_d$ and $M_s$ are the number of data symbols and fixed symbols, respectively. The fixed symbols are transmitted before the data symbols to enable frequency domain equalization (FDE). After up-sampling, filtering, and down-sampling, the $i$th SC block of the SC packet $\mathbf{x}_i \in \mathbb{C}^{T_{tx}\times 1}$ can be represented as

$$\mathbf{x}_i = \sqrt{b} \downarrow_b \mathbf{P}_{tx} \uparrow_a \begin{bmatrix} \mathbf{s} \\ \mathbf{d}_i \end{bmatrix}, \quad (1)$$

where $a \in \mathbb{Z}^+$ is the up-sampling factor, $\uparrow_a \in \mathbb{R}^{K\times M}$ with $M = M_s + M_d$ and $K = a\times M$ is the up-sampling matrix, $\mathbf{P}_{tx} \in \mathbb{C}^{P_{tx}\times K}$ with $P_{tx} = K + L_{tx} - 1$ is the convolution matrix which applies an $L_{tx}$-tap pulse shaping vector denoted by $\mathbf{f} \in \mathbb{C}^{L_{tx}\times 1}$ with $\|\mathbf{f}\|_2 = 1$, and $\downarrow_b \in \mathbb{R}^{T_{tx}\times P_{tx}}$ is the down-sampling matrix, where $b \in \mathbb{Z}^+$ is the down-sampling factor. Without loss of generality, we assume that $b \leq a$ and $T_{tx} = P_{tx}/b = (a\times M + L_{tx} - 1)/b \in \mathbb{Z}^+$, i.e., $P_{tx}$ is an integer multiple of $b$. Typically, $a$ and $b$ are chosen such that the radio operates above the critical sample rate while the choice of $a$ and $b$ may be used to address hardware issues. For example, IEEE 802.11ad [6] suggests $a/b$ to be 3/2 so that both OFDM and SC can operate at the same sample rate.

To model the packet based on the block-based description above, we adopt the overlap-add (OA) method without loss of generality, which is an efficient way to calculate the convolution of a long signal [19]. After the SC blocks are generated via (1), the last $O_{tx}$ samples of the $i$th SC block are overlapped with the first $O_{tx}$ samples of the $(i + 1)$th SC block and summed, as illustrated in Fig. 1a. To maintain the alignment between the overlapping samples of SC blocks and simplify the notation, we assume that $N \triangleq a/b\times M \in \mathbb{Z}^+$, which leads that $O_{tx} = T_{tx} - N = (L_{tx} - 1)/b \in \mathbb{Z}^+$ and $M$ is the integer multiple of $b$.

Let the channel impulse response (CIR) between the transmitter and the receiver be a vector $\mathbf{h} = [h_0\ h_1 \cdots h_{\mathcal{L}}]$ where $\mathcal{L} + 1$ is the number of taps for the multipath channel (MPC). After the SC packet passes through the MPC, we assume that the SC receiver discards the first $N_g$ samples from the received SC packet for the sake of FDE. The receiver then prepares the epochs with the length of $T_{rx} = N + L_{rx}$ for block processing, where $L_{rx}$ is the length of receive filter denoted by $\mathbf{g} \in \mathbb{C}^{L_{rx}\times 1}$. The $i$th SC epoch $\mathbf{e}_i \in \mathbb{C}^{T_{rx}\times 1}$ can be expressed as

$$\mathbf{e}_i = \begin{bmatrix} \mathbf{0}_{T_{rx}\times N_g} & \mathbf{I}_{T_{tx}+\mathcal{L}-N_g} & \mathbf{0}_{N-N_g\times T_{rx}-N+N_g} & \mathbf{0}_{T_{rx}\times \mathcal{L}-q+N-N_g} \\ & \mathbf{0}_{q-\mathcal{L}+N_g\times T_{tx}+\mathcal{L}-N_g} & \mathbf{I}_{T_{rx}-N+N_g} & \end{bmatrix}\begin{bmatrix} \mathbf{y}_i \\ \mathbf{y}_{i+1} \end{bmatrix}, \quad (2)$$

and $\mathbf{y}_l = \mathbf{H}_L \mathbf{x}_i$ where $\mathbf{H}_L \in \mathbb{C}^{(T_{tx}+\mathcal{L})\times T_{tx}}$ is the linear channel convolution matrix, $\mathbf{y}_i$ is the $i$th SC block $\mathbf{x}_i$ alone after passing through the channel, $q = T_{rx} - T_{tx}$. To ensure that $\mathbf{e}_i$ does not contain samples related to $\mathbf{d}_{i-1}$ and $\mathbf{d}_{i+1}$, $N_g$ satisfies the following constraint:

$$O_{tx} + \mathcal{L} \leq N_g \leq a/b \times M_s - O_{rx}, \quad (3)$$

where $O_{rx} = (L_{rx} - 1)/b \in \mathbb{Z}^+$. The upper bound in (3) is because the SC receiver utilizes the first $O_{rx}$ samples of the $i$th SC epoch $\mathbf{e}_i$ for receive filtering. However, this constraint reduces the room for the MPC by $O_{rx}$ samples. To avoid intersymbol interference (ISI), $M_s$ needs to satisfy the condition given by

$$M_s a \geq L_{tx} + L_{rx} + (\mathcal{L}b + 1) - 3. \quad (4)$$

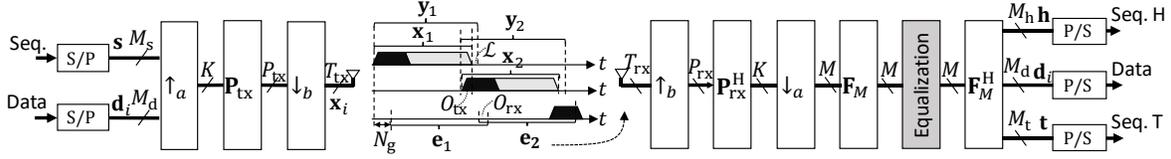

a) SC waveform.

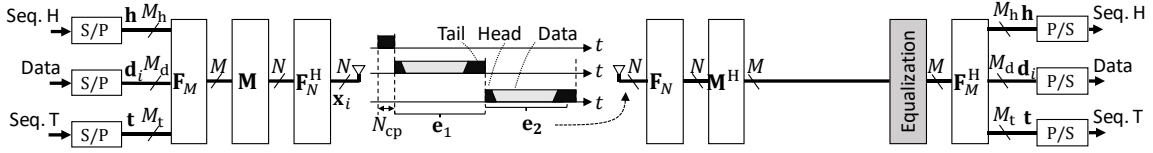

b) UW DFT-s-OFDM.

Fig. 1. The transmit and receive block diagrams for SC waveform with FDE support and UW DFT-s-OFDM.

It is also worth emphasizing that the $i$th SC epoch $\mathbf{e}_i$ is a *cyclically* extended block where the block size is $N$ and the cyclic extension size is $O_{rx}$ samples. This is because the identical fixed sequences are transmitted before and after the data vector $\mathbf{d}_i$. To maintain the same structure for the *last* SC epoch, an additional fixed sequence $\mathbf{s}$ also needs to be transmitted at the end of the packet, as shown in Fig. 1a. This is why IEEE 802.11ad SC PHY attaches an extra Golay sequence to the end of the payload [6].

Based on the discussions above, the overall FDE operations on the $i$th SC epoch can be represented as

$$[\tilde{\mathbf{h}}^H \; \tilde{\mathbf{d}}_i^H \; \tilde{\mathbf{t}}^H]^H = \sqrt{b} \underbrace{\mathbf{F}_M^H \mathbf{E} \mathbf{F}_M}_{\text{FDE}} \downarrow_a \mathbf{P}_{rx}^H \uparrow_b (\mathbf{e}_i + \mathbf{n}_i), \quad (5)$$

where $\mathbf{n}_i \in \mathbb{C}^{T \times 1} \sim \mathcal{CN}(\mathbf{0}_{T_{rx} \times 1}, \sigma_n^2 \mathbf{I}_{T_{rx}})$ is the additive white Gaussian noise (AWGN) with variance $\sigma_n^2$, $\mathbf{F}_M$ is the $M$-point normalized discrete Fourier transform (DFT) matrix, $\mathbf{E}$ is a diagonal matrix for the FDE operation, and $\tilde{\mathbf{h}} \in \mathbb{C}^{M_h \times 1}$ and $\tilde{\mathbf{t}} \in \mathbb{C}^{M_t \times 1}$ are the estimates of the head and tail sequences, i.e., $\mathbf{t}$ and $\mathbf{h}$, respectively, where $\mathbf{s} = [\mathbf{t}^T \; \mathbf{h}^T]^T$ and $M_s = M_h + M_t$. Based on the minimum mean square error (MMSE) criterion, the equalizer matrix $\mathbf{E}$ can be derived as $(\mathbf{Q}^H \mathbf{Q} + \sigma_n^2 \mathbf{I}_M)^{-1}\mathbf{Q}^H$ where

$$\mathbf{Q} = \text{diag}\left\{\sqrt{M}\mathbf{F}_M \downarrow_a \text{circ}\left\{\begin{bmatrix}\mathbf{h}_{\text{effective}}\\\mathbf{0}_{Ma-L_{tx}-L_{rx}+2-b\mathcal{L}-b \times 1}\end{bmatrix}, \tau_0\right\}\right\} \quad (6)$$

in which $\tau_0 = (M_s - M_h)a - N_g b - L_{rx} + 1$ and $\mathbf{h}_{\text{effective}} = \{f\} * \{G\} * \{\uparrow_b \mathbf{h}\}$. Note that $\tau_0$ in (6) is a function of $M_h$ which determines the position of data symbols after the equalization in (5) by circularly shifting the original vector $[\mathbf{s}^T \; \mathbf{d}^T]^T$ by $-M_t$. The variable $M_h$ will be utilized to identify the similarities between SC and UW DFT-s-OFDM in the following sections.

### B. UW DFT-s-OFDM

The block diagram for UW DFT-s-OFDM [10] is given in Fig. 1b. At the transmitter, the data sequence $\mathbf{d}$ and the fixed sequences, i.e., $\mathbf{h}$ and $\mathbf{t}$, are first mapped to the inputs of a DFT matrix $\mathbf{F}_M$. To generate fixed samples at the tail part of each UW DFT-s-OFDM block, the upper-end and the lower-end of DFT-spread, i.e., the first $M_h$ and last $M_t$ columns of $\mathbf{F}_M$, respectively, are allocated for $\mathbf{h}$ and $\mathbf{t}$ since the footprints of $\mathbf{h}$ and $\mathbf{t}$ in time corresponds to the head and tail of each UW DFT-s-OFDM symbol with this allocation, respectively [10]. The output of $\mathbf{F}_M$ is then mapped to a set of $M$ contiguous subcarriers in the FD via a mapping matrix $\mathbf{M} \in \mathbb{R}^{N \times M}$. The spread data and fixed symbols in frequency are converted to the time domain via $\mathbf{F}_N^H$ as

$$\mathbf{x}_i = \mathbf{F}_N^H \mathbf{M} \mathbf{F}_M [\mathbf{h}^H \; \mathbf{d}_i^H \; \mathbf{t}^H]^H, \quad (7)$$

where $\mathbf{x}_i \in \mathbb{C}^{N \times 1}$ is $i$th UW DFT-s-OFDM symbol. $N_{CP}$ samples are added *only* for the first symbol to be able to equalize it with FDE.

Providing that $M_t \geq M/N \times \mathcal{L}$ [18], the $i$th epoch $\mathbf{e}_i$ after the MPC can approximately be represented as

$$\mathbf{e}_i \cong \mathbf{H}_C \mathbf{x}_i, \quad (8)$$

where $\mathbf{H}_C \in \mathbb{C}^{N \times N}$ is a circular channel convolution matrix. The receiver operation can finally be expressed as

$$[\tilde{\mathbf{h}}^H \; \tilde{\mathbf{d}}_i^H \; \tilde{\mathbf{t}}^H]^H = \mathbf{F}_M^H \mathbf{M}^H \mathbf{E} \mathbf{F}_N (\mathbf{e}_i + \mathbf{n}_i), \quad (9)$$

where $\mathbf{n}_i \in \mathbb{C}^{N \times 1} \sim \mathcal{CN}(\mathbf{0}_{N \times 1}, \sigma_n^2 \mathbf{I}_N)$ and $\mathbf{E}$ can be derived based on the MMSE criterion as $(\mathbf{Q}^H \mathbf{Q} + \sigma_n^2 \mathbf{I}_M)^{-1}\mathbf{Q}^H$ and $\mathbf{Q} = \text{diag}\{\sqrt{N}\mathbf{F}_N[\mathbf{h}^T \; \mathbf{0}_{1 \times N-\mathcal{L}-1}]^T\}$.

### III. SINGLE CARRIER VERSUS UW DFT-S-OFDM

In this section, we shed light on the structural differences and similarities, and the performance differences between the SC and UW DFT-s-OFDM waveforms. To this end, we represent the up-sampling and down-sampling operations at the SC transmitter and receiver by using DFT and inverse DFT. We show why SC is suboptimal in MPC as compared to UW DFT-s-OFDM under MMSE FDE. We also discuss the windowing in FD and two-stage equalization for UW DFT-s-OFDM.

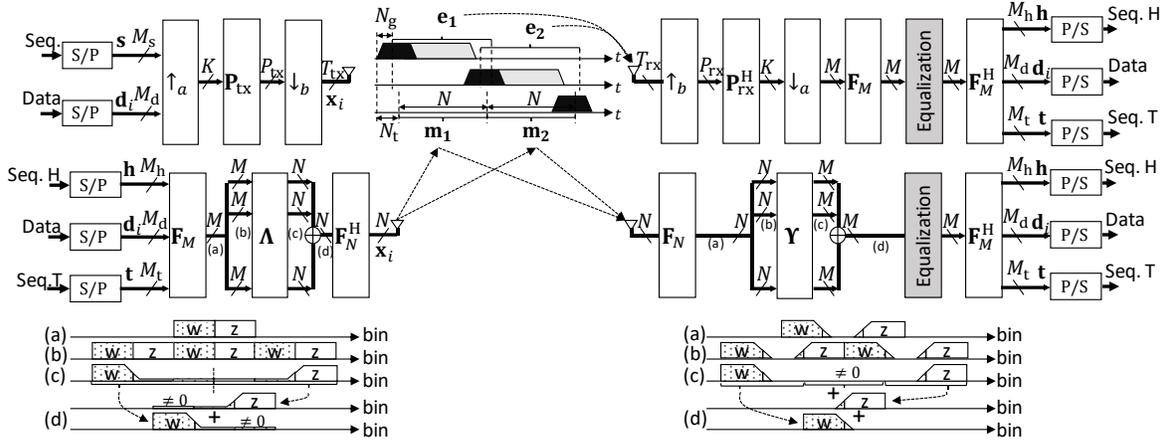

Fig. 2. Equivalent representation of SC waveform with DFT and inverse DFT blocks, and an example for $a = 3$ and $b = 2$.

### A. Up-sampling and Down-sampling at SC Transmitter

Let the consecutive epochs of length $N$ at the output of the SC transmitter after $N_t$ samples from the beginning of the SC packet that are discarded be $\mathbf{m}_i \in \mathbb{C}^{N \times 1}$, where $O_{tx} \leq N_t \leq a/b \times M_s$. The $i$th epoch $\mathbf{m}_i \in \mathbb{C}^{N \times 1}$ can also be calculated as

$$\mathbf{m}_i = \begin{bmatrix} \mathbf{0}_{N \times N_t} & \mathbf{I}_{T_{tx}-N_t} & \mathbf{0}_{N-N_t \times N_t} & \mathbf{0}_{N \times T_{tx}-N_t} \\ \mathbf{0}_{N-T_{tx}+N_t \times T_{tx}-N_t} & & \mathbf{I}_{N_t} & \end{bmatrix} \begin{bmatrix} \mathbf{x}_i \\ \mathbf{x}_{i+1} \end{bmatrix}. \quad (10)$$

Since the last $N_t b$ elements of the corresponding vectors that generate $\mathbf{m}_i$ before linear filtering and down-sampling operations are identical when $O_{tx} \leq N_t \leq a/b \times M_s$, one can express linear pulse shaping as a circular pulse shaping operation. Hence, $\mathbf{m}_i$ can be expressed as

$$\mathbf{m}_i = \sqrt{b} \downarrow_b \mathbf{C}_{tx} \uparrow_a [\mathbf{h}^H \quad \mathbf{d}_i^H \quad \mathbf{t}^H]^H, \quad (11)$$

where $\mathbf{C}_{tx} \in \mathbb{C}^{aM \times aM}$ is a circular convolution matrix where the first column of $\mathbf{C}_{tx}$ is $\text{circ}\{[\mathbf{f}^H \quad \mathbf{0}_{K-L_{tx}+1 \times 1}^T]^T, \tau_1\}$ and $\tau_1 = (M_s - M_h)a - N_t b$. We then exploit the following properties related to the up-sampling and down-sampling operations to represent $\mathbf{m}_i$ with DFT-based operations:

*Property 1 (Up-sampling):* Let $\mathbf{x} \in \mathbb{C}^{M \times 1}$ be a vector of size $M \in \mathbb{Z}$ and let $\mathbf{y} \in \mathbb{C}^{aM \times 1}$ be another vector of size $aM \in \mathbb{Z}$ where $m$th element of $\mathbf{y}$ is equal to $k$th element of $\mathbf{x}$ where $k = (m/a) \in \mathbb{Z}$, otherwise it is 0. Then,

$$\mathbf{y} = \uparrow_a \mathbf{x} = \frac{1}{\sqrt{a}} \mathbf{F}_{aM}^H (\mathbf{1}_{a \times 1} \otimes \mathbf{I}_M) \mathbf{F}_M \mathbf{x}. \quad (12)$$

*Property 2 (Down-sampling):* Let $\mathbf{x} \in \mathbb{C}^{bN \times 1}$ be a vector of size $bN \in \mathbb{Z}$ and let $\mathbf{y} \in \mathbb{C}^{N \times 1}$ be another vector of size $N \in \mathbb{Z}$ where $m$th element of $\mathbf{y}$ is equal to $k$th element of $\mathbf{x}$ where $k = mb \in \mathbb{Z}$. Then,

$$\mathbf{y} = \downarrow_b \mathbf{x} = \frac{1}{\sqrt{b}} \mathbf{F}_N^H (\mathbf{1}_{1 \times b} \otimes \mathbf{I}_N) \mathbf{F}_{bN} \mathbf{x}. \quad (13)$$

The proofs for *Property* 1 and *Property* 2 can be found in [20]. By considering the fact that any circulant matrix can be decomposed as $\mathbf{C}_{tx} = \mathbf{F}_{aM}^H \mathbf{\Lambda} \mathbf{F}_{aM}$ where $\mathbf{\Lambda} = \text{diag}\{\sqrt{aM} \mathbf{F}_{aM} \text{circ}\{[\mathbf{f}^H \quad \mathbf{0}_{Ma-L_{tx}+1 \times 1}^T]^T, \tau_1\}\}$, and employing the identities given in *Property* 1 and *Property* 2, (11) can be rewritten as

$$\mathbf{m}_i = \mathbf{F}_N^H \frac{(\mathbf{1}_{1 \times b} \otimes \mathbf{I}_N) \mathbf{\Lambda} (\mathbf{1}_{a \times 1} \otimes \mathbf{I}_M)}{\sqrt{a}} \mathbf{F}_M [\mathbf{h}^H \quad \mathbf{d}_i^H \quad \mathbf{t}^H]^H \quad (14)$$

By using (14), the following interpretations of the SC transmitter can be made:

*1) Block-based Implementation of SC packet*: (14) shows that the SC packet can be generated by simply concatenating $\{\mathbf{m}_i | i = 1,2,...\}$ synthesized through DFT-based operations. Note that this method is simpler than the overlap-save method which would require $(N + L_{tx} - 1)$-point inverse DFT [19] [20]. However, (14) allows for an $N$-point inverse DFT by exploiting the existence of the UW transmitted before and after $\mathbf{d}_i$.

*2) Difference between SC and UW DFT-s-OFDM Transmitter*: (14) reveals the connection between the SC and UW DFT-s-OFDM waveforms. $\mathbf{\Lambda}$ corresponds to a windowing operation in frequency, where the window is determined by the frequency response of filter $\mathbf{f}$ and which windows the repeated output of the $\mathbf{F}_M$ with a factor of $a$. Since $\mathbf{f}$ is typically a low-pass filter, the filter primarily selects one of the repetitions. As illustrated in Fig. 2, the pass bandwidth of the filter typically is large enough to cover all information (shown as $w$ and $z$ in the illustration, where we consider $N$ frequency bins due to $\mathbf{F}_N^H$) with some roll-off in spectrum. However, since the impulse response of the filter in practice is finite in the time domain, the frequency response of the filter is spread across the bins. On the other hand, if one designs $\mathbf{\Lambda}$ such that it applies a rectangular function, the corresponding transmit diagram in Fig. 2 is identical to the UW DFT-s-OFDM transmitter described in Section II.B. To remove phase rotation in frequency (which also means a symmetric and real filter), the condition on $N_\text{t}$ given by

*Condition 1 (Transmitting SC packet with UW DFT-s-OFDM):*

$$\tau_1 = (M_\text{s} - M_\text{h})a - N_\text{t}b = -(L_\text{tx} - 1)/2 \quad (15)$$

must also hold true. For example, if we consider 802.11ad numerology as $M_\text{s} = 64, a = 3$, and $b = 2$, and assume that $L_\text{tx} = 67$, $N_\text{t}$ must equal to $(225 - 3M_\text{h})/2$ to remove the phase rotation in frequency. When **(15)** is satisfied, the diagonal elements of $\mathbf{\Lambda}$ in **(14)** become real numbers and $\mathbf{m}_i$ is *approximately* a UW DFT-s-OFDM symbol, but not exactly the same due to the additional shaping in frequency. This also implies that the SC receiver can demodulate the UW DFT-s-OFDM packet, but that the data symbols are slightly interfered due to the SC approximation. Note that it is straightforward to limit the support for the diagonal elements of $\mathbf{\Lambda}$ with the FD interpretation of SC. However, it is not possible to achieve to the same $\mathbf{m}_i$ with the time-domain implementation of SC since limiting the support for the diagonal elements of $\mathbf{\Lambda}$ results in a filter length of $Ma$, which causes ISI due to **(4)**.

*3) Interference due to Down-sampling at Transmitter:* **(14)** shows the distortion at the SC transmitter due to down-sampling when $a/b \notin \mathbb{Z}^+$. The operation after shaping in **(14)** divides the resulting vector into $b$ subgroups and overlaps them. The overlapping is not harmful as $a/b \geq 1$, i.e., above the Nyquist rate. However, the frequency response of the filter $\mathbf{f}$ is spread across the bins in practice and the components on the side lobes may still cause interference to the components on the main lobe. Therefore, when $a/b \notin \mathbb{Z}^+$, the receiver may not perfectly recover the symbols with single-tap MMSE FDE due to the small aliasing components even under a noiseless condition. For example, Fig. 2 illustrates the aliasing due to the non-zero components (from $w$ to $z$ or vice versa) on side lobes after down-sampling for $a = 3$ and $b = 2$ in step (d) at the transmitter side. Note that this imperfection can be mitigated with a well-designed filter, and the down-sampling operation can still be considered as not harmful since it reduces the sample rate of the device.

### B. Down-sampling and Up-sampling at SC Receiver

At the receiver side, we exploit the fact that the $i$th SC epoch $\mathbf{e}_i$ is a *cyclically* extended vector by $O_\text{rx}$ samples, which maintains circular convolution of the receive filtering on the last $N$ samples of $i$th SC epoch $\mathbf{e}_i$, to modify **(5)** as

$$[\tilde{\mathbf{h}}^\text{T} \quad \tilde{\mathbf{d}}_i^\text{T} \quad \tilde{\mathbf{t}}^\text{T}]^\text{T} = \sqrt{b}\mathbf{F}_M^\text{H}\mathbf{E}\mathbf{F}_M \downarrow_a \mathbf{C}_\text{rx}^\text{H} \uparrow_b \dot{\mathbf{e}}_i. \quad (16)$$

where $\dot{\mathbf{e}}_i$ is the last $N$ samples of $i$th SC epoch $\mathbf{e}_i$, $\mathbf{C}_\text{rx} \in \mathbb{C}^{aM \times aM}$ is a circular convolution matrix and the first column of $\mathbf{C}_\text{rx}$ is $\text{circ}\left\{[\mathbf{g}^\text{H} \quad \mathbf{0}_{K-L_\text{rx}+1\times1}^\text{T}]^\text{T}, 0\right\}$. Since $\mathbf{C}_\text{rx} = \mathbf{F}_{aM}^\text{H}\mathbf{Y}\mathbf{F}_{aM}$, where $\mathbf{Y} = \text{diag}\left\{\sqrt{aM}\mathbf{F}_{aM}\text{circ}\left\{[\mathbf{g}^\text{H} \quad \mathbf{0}_{Ma-L_\text{rx}+1\times1}^\text{T}]^\text{T}, 0\right\}\right\}$, **(16)** can then be rewritten as

$$[\tilde{\mathbf{h}}^\text{H} \quad \tilde{\mathbf{d}}_i^\text{H} \quad \tilde{\mathbf{t}}^\text{H}]^\text{H} = \mathbf{F}_M^\text{H}\mathbf{E}\frac{(\mathbf{1}_{1\times a}\otimes\mathbf{I}_M)\mathbf{Y}(\mathbf{1}_{b\times 1}\otimes\mathbf{I}_N)}{\sqrt{a}}\mathbf{F}_N\mathbf{m}_i, \quad (17)$$

by using Property 1 and *Property* 2. By inferring **(17)**, the following interpretations on the SC receiver can be made:

*1) Non-coherent additions in Fading Channel:* Assuming that the filter is a Nyquist filter and the filter is known at the receiver side, one may apply a matched filter to the SC epochs. In that case, it is well-known that **(16)** is the optimum receiver in an AWGN channel in the sense that it maximizes SNR. The equivalent operation in FD is that the receiver applies an optimal weight to each bin before changing the dimension of the up-sampled vector of the DFT of $\dot{\mathbf{e}}_i$ before the overlapping operation. On the other hand, **(17)** loses its optimality in a fading channel. Since the DFT of $\dot{\mathbf{e}}_i$ includes the impact of both frequency responses of the transmit filter and MPC via point-to-point multiplications in frequency, the frequency response of the receive filter clearly does not provide the optimal weights as the channel coefficients in frequency are arbitrary complex numbers. Thus, the combination in **(17)**, i.e., $\mathbf{1}_{1\times a}\otimes\mathbf{I}_M$, cause *incoherent* combinations as the phase of each bin is not aligned. For UW DFT-s-OFDM, there is no incoherent combination before equalization due to the perfect rectangular filter in FD. This observation leads to two-stage equalization given in Section III.C when FD windowing applied to UW DFT-s-OFDM.

*2) Decoding SC packet with UW DFT-s-OFDM Receiver:* **(17)** shows the relationship between the SC receiver and the UW DFT-s-OFDM receiver. When a UW DFT-s-OFDM receiver receive an SC block, the UW DFT-s-OFDM receiver equalizes one of the aliases (weighted by the transmit filter) and the discard the rests of the information based on **(9)**. On the other hand, the traditional SC receiver combines the aliases before FDE as in **(17)**. This means that the UW DFT-s-OFDM receiver can receive an SC waveform without any interference, but with some degradation due to discarding other weighted aliases.

### C. Two-stage FDE: First-Phase-Then-Amplitude

The SC waveform yields better PAPR results than the UW DFT-s-OFDM waveform when it employs an FIR filter with high roll-off factor [10]. However, because of the finite support of the filter in time, the SC waveform spreads the information in frequency. On the other hand, the information is localized in the FD with UW DFT-s-OFDM. Therefore, UW DFT-s-OFDM maintains the orthogonality between the frequency-domain resources, which may be important in certain scenarios, e.g., accessing different channels with different beamforming gains in the uplink. To keep the benefit of both SC and UW DFT-s-OFDM, one approach is to apply FD shaping [11] to UW DFT-s-OFDM, so called UW DFT-s-W-OFDM [10]. In this scheme, the support of the diagonal elements of $\mathbf{\Lambda}$ in **(14)** is still limited as in UW DFT-s-OFDM, but it is extended, i.e., larger than $M$, and smoothed with a FD shaping function. This operation modifies the Dirichlet sinc kernel of UW DFT-s-OFDM to reduce the PAPR of UW DFT-s-OFDM. In addition, the UW DFT-s-OFDM packet approximates an SC waveform further with this operation under the condition of **(12)**. To take the energy in the aliases into account, we propose a simple equalization strategy so called two-staged equalization. In the first stage of the equalization, the phase distortion due to the channel is equalized in the FD (i.e., *phase-first*). The aliases are then combined based on their SNRs (note that SNR for each frequency bin differs due to the channel frequency response and the windowing in the FD at transmitter). Since the phases are corrected before the combination, a coherent combination is achieved. In the second stage, the amplitude distortion is equalized before the DFT de-spread operation (i.e., *then amplitude*). Two-stage FDE can also be applied to the SC waveform as long as one considers the FD interpretation of the SC receiver in the implementation. Note that UW DFT-s-OFDM removes the need of prior knowledge of the transmit filter at the receiver side (i.e., the filter is basically a rectangular function in frequency) and a two-stage FDE.

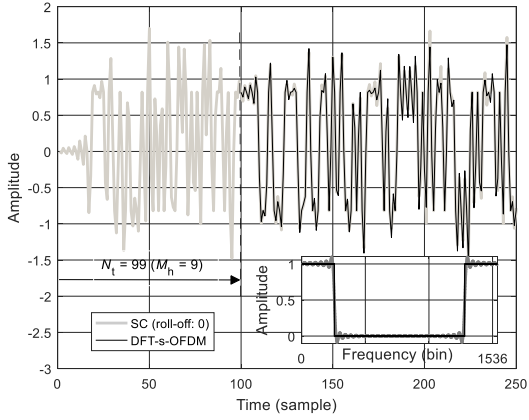

Fig. 3. Temporal characteristics (MSE: -21 dB)

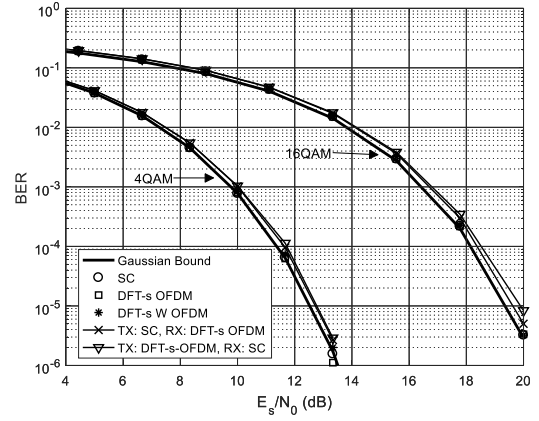

Fig. 5. BER performance in AWGN channel.

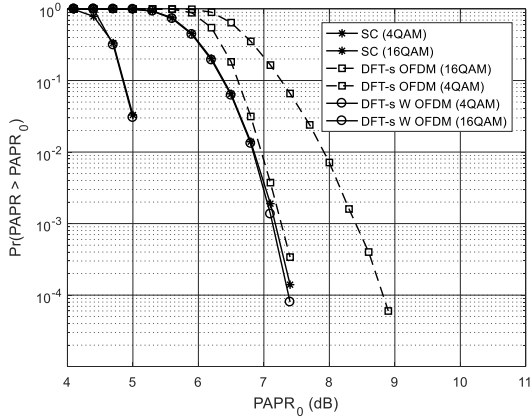

Fig. 4. PAPR performance.

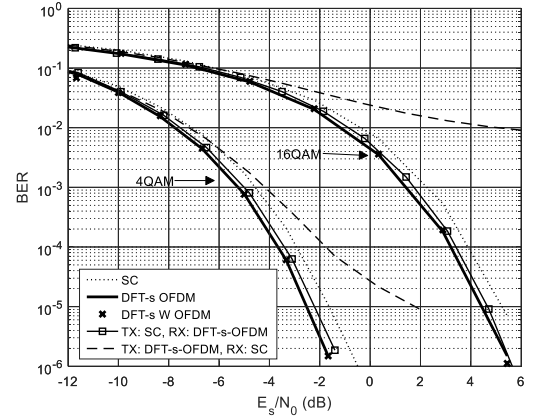

Fig. 6. BER performance in fading channel.

IV. NUMERICAL RESULTS

In this section, we compare the SC and UW DFT-s-OFDM waveforms numerically. We consider the numerology adopted in IEEE 802.11ad for the SC waveform, where $M = 512$, $M_s = 64$, $a = 3$, and $b = 2$. We assume that the SC transmitter employs an $L_{tx} = 67$ tap root-raised cosine (RRC) filter with roll-off factor of $\rho = \{0, 0.2, 0.3\}$ and the receiver uses a matched filter, i.e., $L_{rx} = 67$. For UW DFT-s-OFDM, we assume that $M_h = 9$. For MPC, we consider the IEEE 802.11ad channel model [21], where the scenario is a conference room (STA-AP as subscenario). We assume that the transmitter and receiver employ 8×1 phase antenna arrays (PAAs), unless otherwise stated.

In Fig. 3, we provide temporal characteristics of the UW DFT-s-OFDM and SC waveforms where $\rho = 0$ for BPSK. By using the condition given in **(15)**, we find that $N_t = (225 - 3M_h)/2 = 99$. By offsetting $N_t = 99$ samples from the beginning of the SC packet, we plot the UW DFT-s-OFDM waveform on top of SC waveform. As shown in Fig. 3, the UW DFT-s-OFDM and SC waveforms are approximately the same, where the sample MSE between two waveforms is -21 dB. The difference is due to the imperfect rectangular shape of the SC filter as given in Fig. 3.

In Fig. 4, we provide the PAPR results for QPSK and 16QAM. The SC waveform is 2 dB and 1 dB better than UW DFT-s-OFDM for QSPK and 16QAM, respectively, when $\rho = 0.2$. However, when the FD windowing is applied to UW DFT-s-OFDM, where the amount of the extensions are 51 bins on left and right side of the signal bandwidth and an RRC filter is employed (i.e., corresponds to 0.2 roll-off for UW DFT-s-W-OFDM), the PAPR performance significantly improved and it is almost identical to that of the SC waveform.

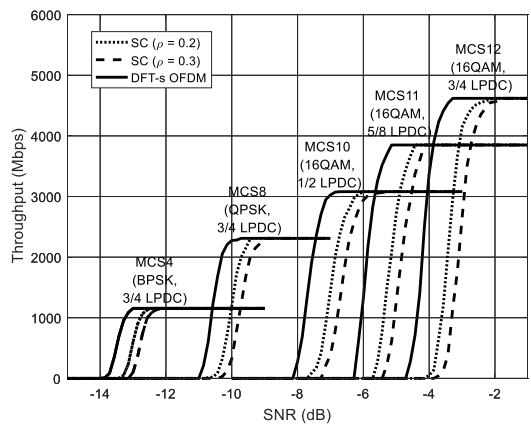

Fig. 7. Throughput with SC and UW DFT-s-OFDM.

In Fig. 5, we investigate the BER performance in AWGN when there is only one antenna element at both transmit and receive PAAs. As expected, UW DFT-s-OFDM, DFT-s-W-OFDM, and SC waveforms achieve the Gaussian BER bound. On the other hand, when there is a mismatch between the transmitter and receiver structures, the BER performance degrades. If the transmitter is for SC and the receiver is for UW DFT-s-OFDM, the performance degrades because of not accounting for the energy on the aliases. When the transmitter is for UW DFT-s-OFDM and the receiver is for SC, the degradation in BER is because the SC receiver cannot perfectly model the Dirichlet sinc kernel.

In Fig. 6 and Fig. 7, we investigate the BER and the peak throughput performance for MCS indices 4, 8, 10, 11, and 12 under an 802.11ad channel model. We consider two-stage FDE for UW DFT-s-W-OFDM. Fig. 6 shows that the SC waveform loses its optimality in a fading channel due to the incoherent additions in FD, as discussed in Section III.B. The performance degradation for SC is about 1 dB, as compared to UW DFT-s-OFDM. Fig. 6 also shows the cross-link performances; while UW DFT-s-OFDM decodes the SC waveform successfully with the degradation of 0.1 dB, SC receiver cannot perfectly decode UW DFT-s-OFDM and suffers from an error floor due to the Dirichlet sinc function as investigated in Section III.A and Section III.B. The peak throughput results in Fig. 7 shows that UW DFT-s-OFDM achieves 1 dB gain as compared to the SC waveform, which indicates that UW DFT-s-OFDM is superior to SC in peak throughput performance in mmWave channels.

V. CONCLUDING REMARKS

In this study, we give insights on the similarities and differences between the SC and UW DFT-s-OFDM waveforms by re-expressing the up-sampling and down-sampling operations of the IEEE 802.11ad SC waveform by using DFT and inverse DFT. This representation reveals that IEEE 802.11ad SC waveform and UW DFT-s-OFDM are structure-wise similar to each other and the IEEE 802.11ad packet structure does not change for UW DFT-s-OFDM. On the other hand, it shows that the SC receiver with down-sampling operation yields suboptimal error rate performance in a fading channel since the down-sampling operation before equalization causes incoherent additions in frequency. The peak throughput results also show that the use of the UW DFT-s-OFDM waveform provides a 1dB gain over the current IEEE 802.11ad SC waveform and the UW DFT-s-OFDM receiver achieves this gain while providing backwards compatibility when receiving IEEE 802.11ad SC waveforms. With cross-link analysis, we show that a UW DFT-s-OFDM receiver can decode an SC waveform with only a slight performance degradation whereas an SC receiver suffers from an error floor due to the Dirichlet Sinc kernel of UW DFT-s-OFDM. We also investigate FD windowing for UW DFT-s-OFDM and two-stage equalization which equalizes the phase and amplitude in separate stages to achieve coherent additions in FD for UW DFT-s-W-OFDM.